\documentclass[12pt,preprint]{aastex}

\usepackage[onecolumn]{emulateapj5}

\shortauthors{White, Becker, Fan, Strauss}
\shorttitle{ACS Observations of \qso}

\slugcomment{Submitted to the Astronomical Journal, 4 November 2004}

\newcommand{\qso}{SDSS J1148$+$5251}
\newcommand{\qa}{SDSS J1030$+$0524}
\newcommand{\fblue}{F720N}
\newcommand{\fred}{F905M}
\newcommand{\lya}{Ly$\,\alpha$}
\newcommand{\lyb}{Ly$\,\beta$}

\newcommand{\taua}{\tau_\alpha}
\newcommand{\taub}{\tau_\beta}
\newcommand{\taubar}{\bar{\tau}}
\newcommand{\erfcx}{\hbox{erfcx}}

\begin{document}
\title{Hubble Advanced Camera for Surveys Observations of the
$z=6.42$ Quasar \qso: A Leak in the Gunn-Peterson Trough\altaffilmark{1}}
\author{
Richard~L.~White\altaffilmark{2},
Robert~H.~Becker\altaffilmark{3,4},
Xiaohui Fan\altaffilmark{5},
\&
Michael A. Strauss\altaffilmark{6}}
\email{rlw@stsci.edu}

\altaffiltext{1}{Based on observations obtained with the NASA/ESA
{\it Hubble Space Telescope} (HST).  HST is operated by the
Association of Universities for Research in Astronomy, Inc. (AURA)
under NASA contract NAS5-26555.}
\altaffiltext{2}{Space Telescope Science Institute, Baltimore, MD 21218}
\altaffiltext{3}{Physics Dept., University of California, Davis, CA 95616}
\altaffiltext{4}{IGPP/Lawrence Livermore National Laboratory}
\altaffiltext{5}{Steward Observatory, University of Arizona, Tucson, AZ 85721}
\altaffiltext{6}{Princeton University Observatory,
Princeton, NJ 08544}

\begin{abstract}

The Hubble Advanced Camera for Surveys has been used to obtain a
narrow-band image of the weak emission peak seen at $\lambda=7205$~\AA\
in the Gunn-Peterson \lyb\ absorption trough of the highest redshift
quasar, \qso.  The emission looks perfectly point-like; there is
no evidence for the intervening galaxy that we previously suggested
might be contaminating the quasar spectrum.  We derive a more
accurate astrometric position for the quasar in the two filters
and see no indication of gravitational lensing.  We conclude that
the light in the \lyb\ trough is leaking through two unusually
transparent, overlapping windows in the IGM absorption, one in the
\lyb\ forest at $z\sim6$ and one in the \lya\ forest at $z\sim5$.

If there are significant optical depth variations on velocity scales
small compared with our spectral resolution ($\sim150$~km/s), the \lya\
trough becomes more transparent for a given \lyb\ optical depth.
Such variations can only strengthen our conclusion that the fraction
of neutral hydrogen in the IGM increases dramatically at $z>6$.  We
argue that the transmission in the \lyb\ trough is not only a more
{\it sensitive} measure of the neutral fraction than is \lya, it
also provides a less {\it biased} estimator of the neutral hydrogen
fraction than does the \lya\ transmission.

\end{abstract}


\section{Introduction}
\label{section-introduction}

The epoch of reionization of intergalactic hydrogen is a key
cosmological observable that constrains the star and galaxy formation
history of the universe.  Clues to the ionization state of the
intergalactic medium (IGM) come from two principal sources: cosmic
microwave background (CMB) polarization measurements, which are
sensitive to the integrated electron scattering optical depth from
$z=0$ to $z\sim1000$, and studies of hydrogen absorption in the
spectra of quasars and galaxies, which can detect neutral gas along
the line of sight at $z \lesssim 6$.

The situation is currently somewhat muddled.  The CMB measurements
from the first year of Wilkinson Microwave Anisotropy Probe (WMAP)
observations indicate that the IGM may already have been ionized
at $z>10$ (Kogut et al.\ 2003).  Quasar spectra show evidence for
a sharp increase in the \lya\ absorption at $z>6$, leading to the
suggestion that reionization occurred near that time (Becker et
al.\ 2001; Fan et al.\ 2002, 2003, 2004; White, Becker, Fan \&
Strauss 2003; but see Songaila 2004 for an opposing view).  Wyithe
\& Loeb (2004) and Mesinger \& Haiman (2004) conclude from the
extragalactic \ion{H}{2} regions around these quasars that the IGM
is indeed close to neutral. On the other hand, the mere detection
of \lya-emitting galaxies at higher redshifts is argued to be
inconsistent with the IGM having a significant neutral fraction at
$z\sim6.5$ (Rhoads et al.\ 2004; Stern et al.\ 2004; Malhotra \&
Rhoads 2004).

Theoretical work has shown that the WMAP and quasar results are
not necessarily contradictory because they are sensitive to rather
different physical regimes (Gnedin 2004).  The black Gunn-Peterson
(1965) \lya\ absorption troughs in quasar spectra can be produced
by a small fraction ($\sim 10^{-2}$) of neutral hydrogen, while
the CMB polarization can be attributed either to an interval at
high redshift when the universe was ionized followed by another
recombination, or to a more recent reionization that persisted to
the present day (see Fig.~7 of Gnedin 2004.) There are also
theoretical reasons to believe that the ionization history of the
universe could have been complex, with multiple periods of ionization
and recombination (Wyithe \& Loeb 2003, Cen 2003).  Recent studies
indicate that even a completely neutral IGM would not necessarily
suppress \lya\ emission from galaxies at $z\sim7$ (Haiman 2002,
Santos 2004, Gnedin \& Prada 2004), especially when such uncertain
effects as galactic winds, clustering of star-forming galaxies,
and radiative transfer in the scattered \lya\ radiation are included.
We need better observations to distinguish among the many possibilities.

\begin{figure*}
\epsscale{0.97}
\plotone{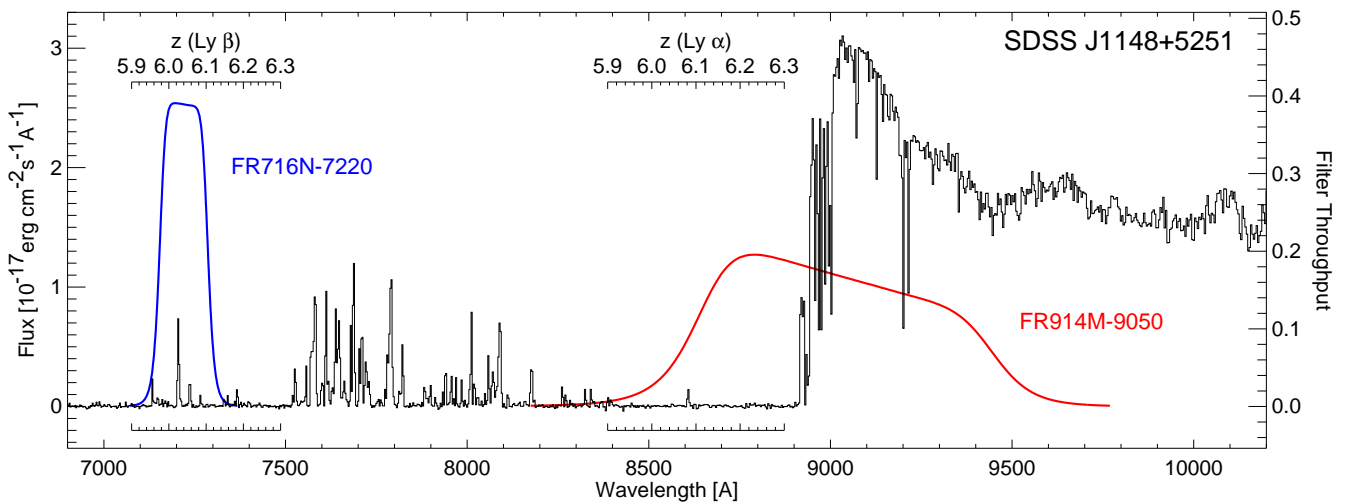}
\caption{
Keck ESI spectrum of \qso\ (White et al.\ 2003)
with the filter throughputs for our HST ACS
observations.  The narrow blue filter (FR716N-7220) is placed in the dark
\lyb\ absorption region around $z=6$ and isolates the narrow emission peak
at $\lambda=7205$~\AA. The broader red filter (FR914M-9050) encompasses
the quasar's strong \lya\ emission line and continuum and is used as
a reference image.
}
\label{fig-spectrum}
\end{figure*}

Our understanding of quasar absorption at $z>6$ is based on only
a handful of quasars, so cosmic variance along the different lines
of sight is clearly a concern.  There are just four
quasars known with redshifts beyond 6.2, all discovered using the
Sloan Digital Sky Survey (Fan et al.\ 2004), and one of those has
intrinsic broad absorption lines that make it unsuitable for studies
of the intervening \lya\ absorption.  White et al.\ (2003) presented
high resolution, high signal-to-noise ratio Keck spectra of the two
highest redshift quasars, SDSSp~J103027.10$+$052455.0 ($z=6.28$,
hereafter \qa) and SDSS~J114816.64$+$525150.3 ($z=6.42$, hereafter
\qso).  \qa\ shows black hydrogen absorption with no indication of
any residual quasar light in either the \lya\ or \lyb\ troughs over
the redshift range $5.97 < z < 6.18$, which sets strong limits on
$\taua$, the \lya\ optical depth derived from the mean transmission.
This spectrum has been analyzed by several different groups (e.g.,
Mesinger \& Haiman 2004) and is the strongest evidence that the
IGM is becoming neutral at $z>6$.

The spectrum of \qso\ also shows deep \lya\ and \lyb\ absorption
but differs from \qa\ in that the absorption troughs are not
completely black (Fig.~\ref{fig-spectrum}).  There is a weak peak
at $z=6.08$ that is seen in both \lya\ and \lyb\ and is likely due
to an ionized bubble along the line-of-sight that produces a ``leak''
in the IGM absorption (White et al.\ 2003).  There are also several
stronger emission peaks in the \lyb\ trough without associated
\lya\ peaks, with the strongest at 7205~\AA.  The straightforward
interpretation of these peaks as due to transparent \lyb\ windows
in the IGM faces some serious problems, however (discussed further
in \S\ref{section-discussion}.) Consequently, White et al.\ suggested
that these features might be \lya\ emission at $z\sim5$ from an
intervening galaxy; the presence of strong \ion{C}{4} absorption
at the same redshift provided some support for this suggestion.
The implausibility of finding such a galaxy along the line-of-sight
by chance would be reduced if the intervening object is lensing
the quasar.  The uncertainty about the nature of the residual light
in the \lya\ and \lyb\ troughs of \qso\ called into question its
use for studying the ionization state of the IGM at $z\sim6$.

In this paper, we report Hubble Space Telescope (HST) Advanced
Camera for Surveys (ACS) observations of the quasar \qso\ using a narrow
filter in the \lyb\ trough centered on the 7205~\AA\ peak.  These
ACS images discriminate between the two hypotheses by allowing us
to look directly for evidence that the emission comes from an
intervening galaxy and to check for any signs that the quasar is
gravitationally lensed.  The observations are described in
\S\ref{section-observations}, and \S\ref{section-analysis} discusses
the analysis.  The results and their implications for this object
and for the IGM ionization are dicussed in \S\ref{section-discussion}.

\section{Observations}
\label{section-observations}

Images of \qso\ were obtained with the ACS Wide Field Camera (WFC)
on 2003 October 25.  Two ramp filters were used: a narrow band
covering the emission peak at 7205~\AA\ (FR716N-7220, 4256~s exposure
time) and a broader band located on the quasar's \lya\ emission
line (FR914M-9050, 720~s).  We briefly refer to these filters
as \fblue\ and \fred\ in the following.  Figure~\ref{fig-spectrum}
shows the location of the filter bandpasses with respect to the
quasar's spectral features.  About 60\% of the \fblue\ counts come
from the narrow feature at 7205~\AA, with most of the remainder
coming from the weaker features nearby.  The short-exposure
\fred\ image is used to determine accurately the position and
morphology of the quasar for comparison with the \fblue\ detection.

Note that the field of view in the ACS ramp filters is limited to
about $40\times80$~arcsec. The filter strip is narrow, covering
only about a third of the ACS field of view, and the ramp central
wavelength changes along the length of the filter so that different
parts of the image are being observed at different wavelengths.
This does complicate the data analysis, but the limited field does
not interfere with our science goals.

The images have been processed using the ACS Science Team's APSIS
pipeline (Blakeslee et al.\ 2003), which aligns the images, corrects
the astrometry to the GSC-2 coordinate system, produces combined
drizzled images on the same output pixel grid for the two filters,
and runs SExtractor (Bertin \& Arnouts 1996) to identify sources in
a detection image created by combining the two filter images.
We have augmented the processing using a custom cosmic-ray (CR)
rejection algorithm that is optimized for the case of only two CR
rejection images.  The algorithm recognizes instances where cosmic
rays hit the same pixel in both images and flags both as bad.  The
observations included two CR-split exposures at each of two different
dither positions for \fblue\ and a single exposure at the two
dither positions for \fred.

\begin{figure*}
\epsscale{0.6}
\plotone{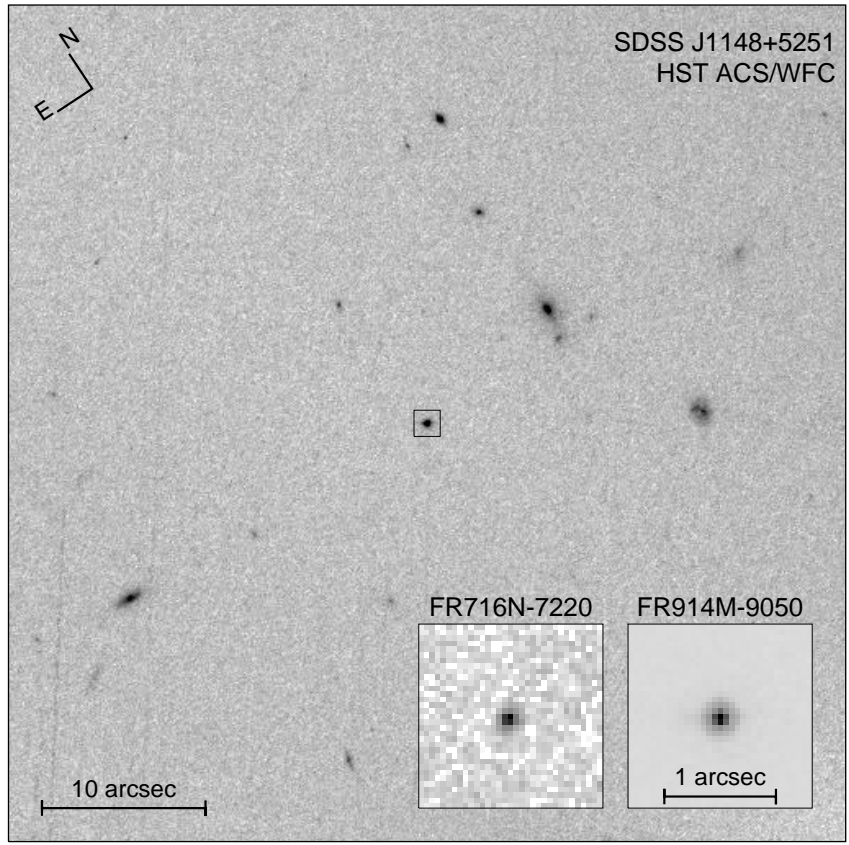}
\caption{
HST ACS/WFC images of \qso.  The large image shows the detection image (a combination
of the two filters), while the insets depict the quasar image in the two filters.
The quasar appears unresolved, and the images coincide to high accuracy.
}
\label{fig-images}
\end{figure*}

Figure~\ref{fig-images} shows the resulting ACS images.  The large
image is the detection image while the insets show the quasar in
the two filters.  The detection image 
is dominated by the long-exposure \fblue\ image
except for the quasar, which is much brighter in the \fred\ image.
The count rate for the quasar is 0.83 counts/s in \fblue\ and 54.5
counts/s in \fred.

An inspection reveals no indication that either the \fblue\ or
\fred\ image is resolved, nor is there a significant shift between
them.  We quantify this statement in the next section, but our
basic conclusion is that the 7205~\AA\ emission peak is perfectly
consistent with quasar light leaking through a transparent window
in the $z\sim6$ \lyb\ and $z\sim5$ \lya\ IGM absorption.

\section{Analysis}
\label{section-analysis}

Our principal goal in acquiring these observations was to determine
the nature of the emission seen in the \lyb\ absorption trough at
7200~\AA.  We therefore focus on determining
whether the emission is extended (as expected for an intervening
galaxy) or point-like (as expected for quasar light) and whether
there is any indication of an astrometric shift between the two
images (which would be expected if the quasar were gravitationally
lensed.) We also determine accurate absolute astrometry for the
quasar using nearby SDSS objects as references.

\subsection{Size Limits}
\label{section-size}

An upper limit on the size of the \fblue\ image was determined by
fitting a monochromatic 7205~\AA\ Tiny Tim (Krist 1995) point-spread
function (PSF) to the \fblue\ quasar image.  The simulated PSF was
subsampled $7\times7$ to allow an accurate sub-pixel position match.
The fitting was performed on the flat-fielded image rather than
the drizzled image to avoid possible problems with the drizzled
PSF and to incorporate the best possible knowledge of the noise
characteristics of the data.  (Fits to the drizzled image gave
similar results.)

The data are fitted very well by the PSF plus sky; the reduced $\chi^2$
for a $20\times20$ pixel region ($1\times1\,\hbox{arcsec}^2$)
around the quasar is less than unity.
To determine the size limit, the PSF was blurred with a Gaussian.
When the Gaussian width is too large, it is no longer possible to
get a good fit as determined by the $\chi^2$ value.  The 90\%
confidence upper limit on the source size is a FWHM of 0.059~arcsec,
and the 99\% confidence limit is 0.067~arcsec.

For comparison, galaxies at $z=5$ typically have half-light diameters
of 0.5~arcsec, and are almost never seen to have diameters smaller
than 0.2~arcsec (Ferguson et al.\ 2004).  Our 99\% confidence size
limit is far smaller than would be expected for an intervening
$z=5$ galaxy.

While we can confidently conclude that most of the light in \fblue\
does not come from a galaxy, we cannot rule out the possibility
that a small fraction of the light is extended.  Fitting the sum
of a point source and a 0.5~arcsec FWHM Gaussian reveals that even
a galaxy with 30\% of the total flux would not be detectable.  The
90\% confidence upper limit on the fraction of light contributed
by a galaxy is 0.46.  That corresponds to a continuum level of
$2.2\times10^{-19}\,\hbox{erg}\,\hbox{cm}^{-2}\,\hbox{s}^{-1}\,\hbox{\AA}^{-1}$.
The best limit on residual continuum emission from an intervening
galaxy probably comes from the spectrum itself and is about
$5\times10^{-20}$.

\subsection{Relative Astrometry}
\label{section-relative-astrometry}

The relative astrometric alignment of the two filter images was
established by comparing positions for all common sources in the
field of view.  We fitted Tiny Tim PSFs to the drizzled images to
determine the quasar position in each filter.  (In this case we
used the drizzled images because we needed to correct for the ACS
distortion.)  The measured quasar positions differ by only 0.22~pixels
(0.011~arcsec).  This is very similar to the rms for the objects
matched to align the images.  We conclude that the images in the
two filters have positions that agree to better than 0.02~arcsec.

Note that if a lensing galaxy produced some of the
light being detected in the \fblue\ filter, we would expect to see
some apparent shift between the positions because the apparent
lensed quasar position would be shifted away from the galaxy center,
and the galaxy would be far more prominent in the \fblue\ filter.
There is no hint of any such effect.  Richards et al.\
(2004) also found no evidence for lensing in HST ACS images of
four other high-redshift quasars.

\subsection{Absolute Astrometry}
\label{section-absolute-astrometry}

We corrected the absolute astrometry of the HST images by matching
our SExtractor catalog with the SDSS DR3 photometric
catalog (Abazajian et al.\ 2004)
and measuring the mean shift between them.  We excluded
the quasar itself from this matched list to avoid biases.  There
are 29 ACS/SDSS matches within 50~arcsec of the quasar.  We searched
for matches with separations as large as 2~arcsec, but all the SDSS
sources found fell within 0.6~arcsec of the (APSIS-corrected) ACS
positions.  We conclude that all the matches are reliable associations.

The coordinate offsets in RA and Dec were estimated using both mean
and median statistics, with essentially identical results.  Both
RA and Dec offsets are accurate to 0.03~arcsec, which is our
remaining systematic error compared with the local SDSS coordinate
system.  Of course, this does not mean that our absolute astrometry
is that good, but the astrometric accuracy for the quasar position
is about 0.1~arcsec, the limit set by the SDSS reference frame
(Pier et al.\ 2003).

The corrected position for the quasar \qso\ is $\alpha=11~48~16.645$,
$\delta=+52~51~50.21$ (J2000) with uncertainties of 0.003~s in RA
and 0.03~arcsec in Dec.  This differs by 0.13~arcsec from the quasar
position given in the DR3 source catalog.  Note that the extreme
colors of this object (it is detected only in the z-band longward
of 9000~\AA) introduce significant differential refraction in the
ground-based SDSS images, which reduces the accuracy of the SDSS
astrometry.  Differential refraction is of course not an issue for
HST observations.

This improved position is most valuable in comparing to observations
at other wavelengths.  Walter et al.\ (2004) resolved the CO emission
associated with \qso\ into two components.  The CO positions are
accurate, but the SDSS position falls between the components with
uncertainties that overlap both.  Walter et al.\ speculated that
the quasar should be aligned with their southern component based
on the overall symmetry of the source. Our improved position confirms
that, with the quasar position falling exactly on top of the southern
CO component.

\section{Discussion}
\label{section-discussion}

Our conclusion from the ACS images is that there is no evidence
for a galaxy along the line-of-sight to \qso.  The emission peak detected
in the Gunn-Peterson \lyb\ trough at 7205~\AA\ is light from the
quasar that is leaking through transparent regions in the IGM.

We recognize that our arguments for an intervening galaxy were met
with some (apparently well deserved) skepticism among our
readers\footnote{And even among the coauthors of this paper.}.
But the superficially more plausible explanation that the light is
coming from the quasar actually faces some serious difficulties
that were discussed in White et al.\ (2003).

The principal difficulty is the absence of \lya\ emission associated
with the 7205~\AA\ \lyb\ emission peak.  The \lya\ emission would
be seen at 8539~\AA\ but is completely absent (see Fig.~5 in White
et al.) The optical depth in \lyb\ is smaller than \lya, so it is
expected to have more light leak through the \lyb\ forest.  However,
the $z\sim6$ \lyb\ trough also suffers from overlying $z\sim5$
\lya\ forest absorption, which reduces the incidence of \lyb-only
emission.  And transparent \lyb\ windows are always expected to
have associated \lya\ peaks.

The combined optical depth\footnote{Here we adopt $z=6.42$ for the
quasar, as derived by Willott, McLure \& Jarvis (2003) and confirmed
by the CO observations of Bertoldi et al.\ (2003).  We also use
a ratio $\taub/\taua = 0.16$, correcting the 15\%\ error pointed
out by Songaila (2004).  The effect of these changes on the computed
optical depths is small.} at 7205~\AA\ is $\taua(z{=}5) + \taub(z{=}6)
= 0.97$.  The absence of any \lya\ emission at 8539~\AA\ implies
a $3\sigma$ lower limit $\taua(8539)>3.3$, corresponding to
$\taub(7205)>0.53$.  We conclude that $\taua(z{=}5) < 0.44$.

As White et al.\ point out, there are very few windows in the $z=5$
\lya\ forest with such low optical depths.  The probability that
a randomly selected wavelength near 7200~\AA\ will fall in such a
window is only about 1\%.  That is the fundamental problem: the
strong peak in the \lyb\ trough requires the alignment of two
unusually transparent windows, one in the \lyb\ forest and one in
the \lya\ forest.  However improbable it appears, this is the only
explanation we see for the observed spectrum.

Here we are using the approach criticized by Songaila (2004), simply
scaling the measured $\taua$ and $\taub$ values by the ratio of
the transition strengths.  This does in fact overestimate the
$\taua/\taub$ ratio when there is a distribution of optical depths
within a flux resolution element.  However, the effect discussed
by Songaila always {\sl reduces} the \lya\ optical depths, so a
simple scaling of $\taub$ gives a conservative lower limit to the
true optical depth.  Suppose that the true \lya\ optical depth in
some region has a distribution $P(\tau)$, normalized such that
$\int P(\tau)d\tau = 1$.  Then the observed mean optical depths in
\lya\ and \lyb\ are:
\begin{equation}
\taua = -\ln\left[ \int_0^\infty P(\tau)\exp(-\tau)d\tau \right] \quad,
\label{eqn-taua}
\end{equation}
and
\begin{equation}
\taub = -\ln\left[ \int_0^\infty P(\tau)\exp(-f\tau)d\tau \right] \quad,
\label{eqn-taub}
\end{equation}
where $f=0.16$ is the nominal ratio between the \lya\ and \lyb\ optical
depths.

The exact value of the optical depth ratio depends on the distribution
$P(\tau)$, which has been explored using physical models of the gas
density distribution and ionization balance (Fan et al.\ 2002, Lidz
et al.\ 2002, Cen \& McDonald 2002, and references therein).  For
this paper we prefer to keep the equation in terms of the optical depth
distribution $P(\tau)$ rather rely on a more complicated physical
model of the IGM to clarify the connections to directly observable
quantities.  A simple $P(\tau)$ example illustrates the size of
the effect.  If $P(\tau)$ is a Gaussian with mean $\taubar$ and
width $\sigma$,
\begin{equation}
\taua = \ln\left\lbrace \erfcx[-\taubar/\sqrt{2}\sigma] \over
\erfcx[(\sigma^2-\taubar)/\sqrt{2}\sigma]
\right\rbrace \quad ,
\label{eqn-gausstau}
\end{equation}
where $\erfcx(x)$ is the scaled complementary error function,
\begin{equation}
\erfcx(x) = {2\over\sqrt{\pi}} \int_x^\infty e^{x^2-t^2} dt \quad.
\label{eqn-erfcx}
\end{equation}
When $\taubar \gg \sigma^2$,
\begin{equation}
\taua \simeq \taubar - {\sigma^2 \over 2} \quad .
\label{eqn-gausstaua}
\end{equation}
The equation for $\taub$ is identical but with
$\taubar$ and $\sigma$ replaced by $f\taubar$ and $f\sigma$,
respectively.

The optical depth distribution effect explains the overall ratio
of the mean transmissions observed in the \lya\ and \lyb\ Gunn-Peterson
troughs for \qso. Using the spectrum in Figure~\ref{fig-spectrum}
(White et al.\ 2003), the mean optical depth in the \lyb\ trough
($6.1<z<6.32$) is $\taub = 3.1\pm0.1$ while the mean in the \lya\
trough is $\taua = 6.4\pm0.3$, which is smaller by a factor of three
than might be expected from the optical depth ratio $f=0.16$. (Note
that in both absorption troughs in \qso\ we detect residual light,
whereas in \qa\ we measure only an upper limit to the IGM transmission.)
The transparency of the \lya\ trough is due to the optical depth
distribution in the trough.  If we apply the Gaussian model
(eqn.~\ref{eqn-gausstau}) we find $\taubar = 26.4$ and $\sigma =
10$.  Of course the Gaussian model used for this analysis is not
particularly realistic, but the results will be similar for other
$P(\tau)$ models.

In this case the $\taub/\taua$ ratio is 0.48, which is (as demanded
by the observations) indeed three times larger than $f=0.16$.  But
the change is almost entirely the result of a reduction in the
effective $\taua$ compared with the true mean optical depth $\taubar$.
This effect is easy to understand:  $\taua$ begins to saturate when
the high optical depth regions effectively transmit no light, and
the mean transmission becomes dominated by light in low opacity
regions.

This is an important point: an optical depth distribution makes the
IGM {\sl more transparent} in the \lya\ absorption trough.  As the
\lya\ optical depth rises with increasing redshift, this effect
tends to flatten the $d\taua/dz$ slope.  However, we find that
$\taua$ increases more steeply with $z$ starting at $z\sim6$;
correcting for the $P(\tau)$ effect makes this change even more
dramatic.  The inclusion of this effect thus strengthens the evidence
for a significant increase in the neutral hydrogen fraction beyond
redshift 6.  This is inconsistent with Songaila's (2004) conclusion
that the inclusion of an optical depth distribution weakens the
argument for an abrupt change in the ionization level at $z>6$.

Another way of stating this conclusion is that $\taub/f$
is actually a less biased estimator of the mean neutral hydrogen
fraction (the physical quantity of interest) than is $\taua$.
The interpretation of the observed \lyb\ transmission is far less
reliant on the details of $P(\tau)$ than is the interpretation of
the \lya\ transmission.  Small changes in $P(\tau)$ at $\tau\lesssim1$
can have a large effect on $\taua$, but $\taub$
is sensitive only to $P(\tau)$ changes integrated over the much
larger range $\tau \lesssim 1/f \approx 6$.  The observed $\taub$
optical depths increase very dramatically with redshift for the
known $z>6$ quasars (White et al.\ 2003; Fan et al.\ 2004).  We
interpret this as powerful direct evidence in favor of a rapid
increase in the hydrogen neutral fraction at those redshifts.

Returning to the question of the 7205~\AA\ emission in the \lyb\
trough of \qso, it is clear that that the $P(\tau)$ effect makes
the absence of an associated \lya\ peak even more difficult to
understand because it increases the strength expected for the \lya\
peak.  For example, the specific Gaussian $P(\tau)$ described above
($\taubar/\sigma=2.6$) is marginally consistent with the absence
of \lya\ at 8539~\AA\ only if there is practically no $z=5$ \lya\
forest absorption at all, $\taua(z{=}5) < 0.25$.  It is possible
to rule out $P(\tau)$ distributions with stronger tails near $\tau=0$.
The properties of this system can consequently put limits on the
permitted optical depth distribution $P(\tau)$ at velocity resolutions
finer than those directly observed.

Perhaps in this quasar the $z=6.025$ IGM is very transparent in
\lyb\ yet opaque in \lya\ because the damping wings of the (roughly)
neutral gas on either side are optically thick in \lya\ but not in
\lyb.  If the presence or absence of damping wings can be shown to
play a crucial role in interpreting the \lyb\ ``leaks'' in this
system, then we may be able to strengthen the case for (or against)
a substantial neutral hydrogen fraction in the IGM at $z>6$.

There is at least one satisfying aspect of this ``two window'' model
for the 7205~\AA\ peak.  A plausible way to create a transparent
region in the \lya\ forest is for a star-forming galaxy close to
the line of sight at $z=4.925$ to ionize a bubble in the \lya\
forest.  The halo of that same galaxy probably hosts the gas cloud
that produces the strong \ion{C}{4} absorption doublet seen at
9200~\AA\ ($z=4.943$, White et al.\ 2003).  Note that our observations
do not rule out a faint galaxy roughly coincident with the quasar
(\S\ref{section-size}) even though the light in the 7205~\AA\
emission peak certainly comes from the quasar and not a galaxy.
The angular scale at $z=4.9$ is $6.3$~kpc/arcsec (using the WMAP
cosmology), so the galaxy
need not lie too close to the line of sight to the quasar.  A single
unusual object along the line of sight can thus explain both the
hole in the \lya\ forest absorption and the strong high-ionization
metal absorption.

Finally, since the \lya\ and \lyb\ troughs of \qso\ do show a low level
of detectable light (which is perhaps still slightly contaminated by
the galaxy), this spectrum is worthy of additional detailed physical
modeling as has been applied to \qa.  And given the apparently
unusual characteristics of this line of sight, additional high
redshift quasars sampling the IGM absorption in other directions would
of course be of great value.

\acknowledgments

Thanks to John Blakeslee for advice on some aspects of the data
analysis.
Support for this work was provided by NASA through grant
HST-GO-09734.01-A from the Space Telescope Science Institute, which
is operated by the Association of Universities for Research in
Astronomy, under NASA contract NAS5-26555.
RLW thanks Holland Ford and the Johns Hopkins University Astronomy
\& Physics Department for their hospitality during a sabbatical,
when this work was proposed and the initial analysis was begun.
RHB acknowledges support from NSF grant AST-00-98355 and the
Institute of Geophysics and Planetary Physics (operated under the
auspices of the U.~S.\ Department of Energy by University of
California, Lawrence Livermore National Laboratory under contract
No.~W-7405-Eng-48).
XF acknowledges support from the University of Arizona and an Alfred
P.~Sloan Research Fellowship.
MAS acknowledges support from NSF grant AST-03-07409.


\begin{thebibliography}{}

\bibitem[Abazajian et al.(2004)]{2004astro.ph..10239A} Abazajian, K., et 
al.\ 2004, ArXiv Astrophysics e-prints, astro-ph/0410239 

\bibitem[Becker et al.(2001)]{2001AJ....122.2850B} Becker, R.~H.~et al.\ 
2001, \aj, 122, 2850 

\bibitem[Bertin \& Arnouts(1996)]{1996A&AS..117..393B} Bertin, E.~\& 
Arnouts, S.\ 1996, \aaps, 117, 393 

\bibitem[Bertoldi et al.(2003)]{2003A&A...409L..47B} Bertoldi, F., et al.\ 
2003, \aap, 409, L47 

\bibitem[Blakeslee et al.(2003)]{2003adass..12..257B} Blakeslee, J.~P., 
Anderson, K.~R., Meurer, G.~R., Ben{\'{\i}}tez, N., \& Magee, D.\ 2003, ASP 
Conf.~Ser.~295: Astronomical Data Analysis Software and Systems XII, 12, 
257 

\bibitem[Cen \& McDonald(2002)]{2002ApJ...570..457C} Cen, R.~\& McDonald, 
P.\ 2002, \apj, 570, 457 


\bibitem[Cen(2003)]{2003ApJ...591...12C} Cen, R.\ 2003, \apj, 591, 12 

\bibitem[Fan et al.(2002)]{2002AJ....123.1247F} Fan, X., Narayanan, V.~K., 
Strauss, M.~A., White, R.~L., Becker, R.~H., Pentericci, L., \& Rix, H.\ 
2002, \aj, 123, 1247 

\bibitem[Fan et al.(2003)]{2003AJ....125.1649F} Fan, X., et al.\ 2003, \aj, 
125, 1649 

\bibitem[Fan et al.(2004)]{2004AJ....128..515F} Fan, X., et al.\ 2004, \aj, 
128, 515

\bibitem[Ferguson et al.(2004)]{2004ApJ...600L.107F} Ferguson, H.~C., et 
al.\ 2004, \apjl, 600, L107 

\bibitem[Gnedin(2004)]{2004ApJ...610....9G} Gnedin, N.~Y.\ 2004, \apj, 610, 9 

\bibitem[Gnedin \& Prada(2004)]{2004ApJ...608L..77G} Gnedin, N.~Y.~\&
Prada, F.\ 2004, \apjl, 608, L77 

\bibitem[Gunn \& Peterson(1965)]{1965ApJ...142.1633G} Gunn, J.~E.~\& 
Peterson, B.~A.\ 1965, \apj, 142, 1633 

\bibitem[Haiman(2002)]{2002ApJ...576L...1H} Haiman, Z.\ 2002, \apjl, 576,
L1

\bibitem[Kogut et al.(2003)]{2003ApJS..148..161K}
Kogut, A., Spergel, D. N., Barnes, C., Bennett, C. L., Halpern, M.,
Hinshaw, G., Jarosik, N., Limon, M., Meyer, S. S.,
Page, L., Tucker, G. S., Wollack, E., \& Wright, E. L.
2003, \apjs, 148, 161 

\bibitem[Krist(1995)]{1995adass...4..349K} Krist, J.\ 1995, ASP Conf.~Ser.~ 
77: Astronomical Data Analysis Software and Systems IV, 4, 349 

\bibitem[Lidz, Hui, Zaldarriaga, \& Scoccimarro(2002)]{2002ApJ...579..491L} 
Lidz, A., Hui, L., Zaldarriaga, M., \& Scoccimarro, R.\ 2002, \apj, 579, 
491 

\bibitem[Malhotra \& Rhoads(2004)]{2004astro.ph..7408M} Malhotra, S.~\& 
Rhoads, J.\ 2004, ArXiv Astrophysics e-prints, astro-ph/0407408 

\bibitem[Mesinger \& Haiman(2004)]{2004ApJ...611L..69M} Mesinger, A.~\& 
Haiman, Z.\ 2004, \apjl, 611, L69 

\bibitem[Pier et al.(2003)]{2003AJ....125.1559P} Pier, J.~R., Munn, J.~A., 
Hindsley, R.~B., Hennessy, G.~S., Kent, S.~M., Lupton, R.~H., \& Ivezi{\' 
c}, {\v Z}.\ 2003, \aj, 125, 1559 

\bibitem[Rhoads et al.(2004)]{2004ApJ...611...59R} Rhoads, J.~E., et al.\ 
2004, \apj, 611, 59 

\bibitem[Richards et al.(2004)]{2004AJ....127.1305R} Richards, G.~T., et 
al.\ 2004, \aj, 127, 1305

\bibitem[Santos(2004)]{2004MNRAS.349.1137S} Santos, M.~R.\ 2004, \mnras,
349, 1137

\bibitem[Songaila(2004)]{2004AJ....127.2598S} Songaila, A.\ 2004, \aj, 127, 
2598 

\bibitem[Stern et al.(2004)]{2004astro.ph..7409S} Stern, D., Yost, S.~A., 
Eckart, M.~E., Harrison, F.~A., Helfand, D.~J., Djorgovski, S.~G., 
Malhotra, S., \& Rhoads, J.~E.\ 2004, ArXiv Astrophysics e-prints, 
astro-ph/0407409 

\bibitem[Walter et al.(2004)]{2004ApJ...615L..17W} Walter, F., Carilli, C., 
Bertoldi, F., Menten, K., Cox, P., Lo, K.~Y., Fan, X., \& Strauss, M.~A.\ 
2004, \apjl, 615, L17 

\bibitem[White, Becker, Fan, \& Strauss(2003)]{2003AJ....126....1W} White, 
R.~L., Becker, R.~H., Fan, X., \& Strauss, M.~A.\ 2003, \aj, 126, 1

\bibitem[Willott, McLure, \& Jarvis(2003)]{2003ApJ...587L..15W} Willott, 
C.~J., McLure, R.~J., \& Jarvis, M.~J.\ 2003, \apjl, 587, L15 

\bibitem[Wyithe \& Loeb(2003)]{2003ApJ...586..693W} Wyithe, J.~S.~B.~\& 
Loeb, A.\ 2003, \apj, 586, 693

\bibitem[Wyithe \& Loeb(2004)]{2004Natur.427..815W} Wyithe, J.~S.~B.~\& 
Loeb, A.\ 2004, \nat, 427, 815

\end{thebibliography}
\end{document}